\documentclass{mem}
\usepackage[T1]{fontenc}
\usepackage[utf8]{inputenc} 
\usepackage{natbib}
\usepackage{txfonts}
\usepackage{balance}
\usepackage{flushend}
\usepackage{graphicx}
\idline{75}{282}
\begin{document}
\def\teff{$T\rm_{eff }$}
\def\kms{$\mathrm {km s}^{-1}$}

\title{On the Expected Orbitally-modulated\\ TeV Signatures of Spider Binaries:}
\subtitle{The Effect of Intrabinary Shock Geometry}

\author{
C. \,Venter\inst{1}, A. \, Kopp\inst{1},  Z.\, Wadiasingh\inst{2,3,4} \textit{et al.}\\
The remaining authors can be found at the end of the paper.
          }

\institute{
Centre for Space Research, North-West University, Private Bag X6001, Potchefstroom, South Africa, 2520
\and
Department of Astronomy, University of Maryland, College Park, Maryland 20742, USA
\and
Astrophysics Science Division, NASA Goddard Space Flight Center, Greenbelt, MD 20771, USA
\and
Center for Research and Exploration in Space Science and Technology, NASA/GSFC, Greenbelt, Maryland 20771, USA
\\
The remaining affiliations can be found at the end of the paper.
}

\authorrunning{Venter et al.}

\titlerunning{Spider Binary Shocks}

\date{Received: 20-12-2024; Accepted: XX-XX-XXXX (Day-Month-Year)}

\abstract{
‘Spider’ binary systems – black widow and redback compact binaries differentiated by their companion’s mass and nature – are an important type of pulsar system exhibiting a rich empirical phenomenology, including radio eclipses, optical light curves from a heated companion, as well as non-thermal X-ray and GeV orbital light curves and spectra. Multi-wavelength observations have now resulted in the detection of $\gtrsim50$ of these systems in which a millisecond pulsar heats and ablates its low-mass companion via its intense pulsar wind. Broadband observations have established the presence of relativistic leptons that have been accelerated in the pulsar magnetosphere and near the intrabinary shock, as well as a hot companion, presenting an ideal environment for the creation of orbitally-modulated inverse Compton fluxes that should be within reach of current and future Cherenkov telescopes. We have  included an updated synchrotron kernel, different parametric injection spectral shapes, and several intrabinary shock geometries in our emission code to improve our predictions of the expected TeV signatures from spider binaries. Our updated phase-dependent spectral and energy-dependent light curve outputs may aid in constraining particle energetics, wind properties, shock geometry, and system inclination of several spider binaries.}
\maketitle{}

\section{Introduction}
The collective name `spider binary systems' encapsulates black widow and redback compact binaries. These binaries include as one of their members a millisecond pulsar and they are primarily distinguished by the companion’s mass \citep{Roberts2013}: redbacks have companions with masses $\gtrsim 0.1M_\odot$ while black widows' companion masses are $\ll 0.1M_\odot$. These systems have nearly circular member orbits with orbital periods that are typically less than a day. The collision of pulsar and companion winds or magnetospheres leads to the formation of an intrabinary shock, with Doppler boosting of the radiation by the plasma flowing relativistically along the shock tangent \citep{Arons1993,Tavani1997,Wadiasingh15,Wadiasingh17,Romani16,Kandel2019,vdMerwe2020}. The recent detection of 20~TeV pulsations from the Vela pulsar \citep{Aharonian2023} confirms the presence of multi-TeV particles in pulsar systems. This, together with a heated, tidally-locked companion surface providing a target photon field, may yield orbitally-modulated TeV emission that may be observable for imaging atmospheric Cherenkov telescopes. Moreover, these systems have well-constrained masses and inclination angles due to multi-band observations.

Around 50 spider binaries are now known in the gamma-ray and other bands \citep{Smith2023}; more if one adds new discoveries in the radio band (e.g., \citealt{FAST2020}). The X-ray emission tends to be harder and more luminous in redbacks (where the shock typically wraps around the pulsar), and they are also more efficient emitters (having a larger ratio of X-ray luminosity to spin-down power; \citealt{Swihart2022}) than black widows. Recently, a transitioning millisecond pulsar,
XSS J12270-4859 (4FGL J1228.0-4853) that changed from a low-mass X-ray binary (LMXB) state to a millisecond pulsar state in late 2012, was seen to modulate at the orbital period in the 60~MeV -- 1~GeV band, with a minimum at pulsar inferior conjunction \citep{An22}.
Similar modulation was seen in the pre- and post-transition epochs of this pulsar. \citet{Sim24} modelled
XSS J12270-4859, PSR J2039-5617, and PSR J2339-0533 that exhibited orbital modulation in the X-ray and GeV bands, invoking electron-positron pairs with Lorentz factors of $\gamma\sim10^8$ that pass through the intrabinary shock and penetrate the companion magnetosphere to yield enhanced synchrotron radiation (SR) at the observed orbital phase. For a recent review, see \citet{Sim24_rev}. It is hoped that TeV modulations will also be detected from spider binaries soon.

Phenomenological models of spider binaries include those by \citet{Wadiasingh17,Kim22}, while the intrabinary shock stability was considered by \citet{Wadiasingh18}. Previously \citep{vdMerwe2020}, we constructed an emission model to describe the broadband spectral energy distributions (SEDs) as well as energy-dependent light curves of a few spider binaries, invoking both SR and inverse Compton (IC) scattering by the downstream plasma. \citet{Sullivan23} modelled the expected X-ray polarisation signatures from spiders by considering two magnetic field configurations: toroidal fields and tangential fields that follow the postshock flow. Future polarisation observations may probe the magnetic field structure around the shock. 
Finally, \citep{Cortes22,Cortes24} performed global, 2D particle-in-cell (PIC) simulations of the intrabinary shock in spider binaries, assuming that it wraps around the companion star. They find that the magnetic energy of the relativistic pulsar wind is efficiently converted into particle kinetic energy via shock-driven reconnection. Moreover, Doppler-boosted SR leads to spectra and light curves that resemble the data. 

In this paper, we report some updates to the results  of \citet{vdMerwe2020}, including new parametric forms of the particle injection spectrum, as well as implementing new intrabinary shock structures. Following a brief summary of the spider code (Section~\ref{sec:summary}), we describe some model improvements (Section~\ref{sec:improve}) and the effects thereof on the model outputs, ending with conclusions in Section~\ref{sec:concl}.

\section{Summary of the \texttt{UMBRELA} Code}\label{sec:summary}
We previously described the implementation of an emission code to model the radiative output by spider binaries \citep{vdMerwe2020}. The first part of the code implemented the injection of particles by the pulsar and their transport\footnote{We assume a constant $B$-field along the shock, but as a free parameter.} along the hemispherical intrabinary shock, taking into account SR, IC and adiabatic losses, diffusion, and convection. The SR and IC spectra were calculated in the co-moving frame of the plasma. A parametric bulk velocity that scales linearly with co-latitude was assumed. The second part of the code implemented beaming of emission towards the observer, yielding orbital phase-dependent spectra as well as energy-dependent light curves.

\section{Model Improvements}\label{sec:improve}
\subsection{SR Kernel}
As in \citet{Kopp2013}, we have now updated our SR calculation using the method of \citet{Macleod2000} involving Chebychev polynomials, instead of interpolating the integral of the modified Bessel function $K_{5/3}(z)$ via a table that is logarithmic in the argument $z = E_\gamma/E_{\rm SR}$, with $E_{\rm SR}$ the spectral cutoff energy. The effect is not very pronounced, but the SR spectrum is slightly more accurate and smoother.

\subsection{Injection Spectrum}
We have now included more parametric forms for the injection spectrum. First, we include a power law
\begin{equation}
    Q = Q_0E^{-\Gamma_1}.
\end{equation}
We also considered a broken power law
\begin{equation}
     Q  = Q_0\left(\frac{E}{E_{\rm b}}\right)^{-\Gamma_1}\left(1+\left(\frac{E}{E_{\rm b}}\right)^{(\Gamma_1-\Gamma_2)/f}\right)^{f},
\end{equation}
with $f$ a smoothing factor, and $\Gamma_1$ and $\Gamma_2$ the indices of the respective distribution tails. One has to ensure that $(\Gamma_1-\Gamma_2)/f>0$ to obtain the desired spectral shape.  This form can in some cases mimic the influences of radiative cooling of pairs.
Lastly, we also considered an exponentially cut-off power law
\begin{equation}
    Q = Q_0E^{-\Gamma_1}\exp\left[-\left(\frac{E}{E_{\rm c}}\right)^b\right].
\end{equation}
The strength of the cutoff is set by the parameter $b$, with $b<1$ realising a sub-exponential, $b=1$ an exponential, and $b>1$ a super-exponential cutoff. The normalisation $Q_0$ and minimum particle energy $E_{\rm min}$ (for an otherwise constrained $E_{\rm max}$ and fixed slope(s)) are found by solving the system of equations involving the current and luminosity \citep{Venter15}:
\begin{eqnarray}
    \int_{E_{\rm min}}^{E_{\rm max}}Q\,dE & = & (M_{\rm pair}+1)I_{\rm PC},\\
    \int_{E_{\rm min}}^{E_{\rm max}}EQ\,dE & = & \eta_{\rm part}\dot{E}.
\end{eqnarray}
Here, $M_{\rm pair}$ is the pair multiplicity, $I_{\rm PC}$ the (primary) Goldreich-Julian current from the polar cap, and $\eta_{\rm part}$ is the conversion efficiency of spin-down luminosity $\dot{E}$ to particle acceleration (power). A break or cutoff in the injection spectrum is reflected in the SED, as expected.


\subsection{Extension of and Interpolation on the Photon Energy Grid}
We use a logarithmic energy grid:
\begin{equation}
    E_j = E_0e^{j\kappa},\quad j = 0,1,\dots,N-1,
\end{equation}
with $\kappa = \ln(E_{N-1}/E_0)/(N-1)$.
Upon calculation of the energy of the photon boosted from the co-moving frame to the lab frame ($E = \delta E^\prime$), one has to locate this energy on a grid. The simplest is to obtain the relevant bin index 
\begin{equation}
    j = \frac{\lfloor\ln(\delta E^\prime/E_0)\rfloor}{\kappa},
\end{equation}
with $\delta=\Gamma^{-1}(1-\beta\mathbf{n}\cdot\mathbf{u})^{-1}$ the Doppler factor, $\beta=v/c$, $\mathbf{n}$ the observer direction, and $\mathbf{u}$ the velocity tangent. However, this may lead to some flux not being located at the correct energy bin. We have thus extended this grid, and moreover we interpolated the flux to yield a smoother photon spectrum in the lab frame.

\subsection{Updated Shock Geometry}
The code now includes the following options for the shock geometry: 
\begin{itemize}
    \item The hemispherical geometry that we used in \citet{vdMerwe2020} to approximate the shock geometry near the nose or stand-off distance (labelled `Sphere'). For such a spherical cap, $R(\theta)=R_0$, with $\theta=0$ along the line joining the two stars.
    \item A hemisphere up to a given angle, and then a tangent / long conical tail beyond this angle (labelled `Sphere+Cone'). 
    \item A Wilkin shock \citep{Wilkin96}  due to the interaction of an isotropic stellar wind with a plane-parallel ambient wind. The intrabinary shock radius is given by
\begin{equation}
    R(\theta) = R_0\csc\theta\sqrt{3\left(1-\theta\cot\theta\right)},
\end{equation}
with $R_0$ the stand-off distance.
    \item A Cant\'o shock \citep{Canto96} due to the interaction of two isotropic stellar winds, the geometry depending on $\beta^\prime$, the ratio of the mass-loading rates (momentum fluxes): 
\begin{eqnarray}
    R(\theta) & = &a\sin\theta_1\csc(\theta+\theta_1),\\
    \theta_1\cot\theta_1 & = & 1 + \beta^\prime(\theta\cot\theta-1),\\ 
    \beta^\prime & = & \frac{\dot{M}_1v_1}{\dot{M}_2v_2},
\end{eqnarray}
with $a$ the intrabinary distance. We find $\theta_1$ for a fixed value of $\theta$ by employing the Newton-Raphson method. The stagnation-point radius is given by
\begin{equation}
    R_0 = \frac{(\beta^\prime)^{1/2}a}{1+(\beta^\prime)^{1/2}.}
\end{equation}
In some cases, the shock wraps around the pulsar, while in others the shock wraps around the companion, depending on the strengths of the respective stellar winds. The  shape of the `inverse shock' for $\beta^\prime>1$ can be obtained\footnote{For technical reasons, the maximum value of $\theta$ decreases with $\beta^\prime$.} using $\beta^{\prime\prime} = 1/\beta^\prime$,  providing a mirror image of the first.
\end{itemize}
As a first approximation, we use a parametric expression for the spatial distribution of the particles, with $s$ the arclength along the shock:
\begin{equation}
N_{\rm e} = N_{\rm e,0}\cos\left(\left(\frac{s}{s_{\rm max}}\right)\theta_{\rm max}\right).
\end{equation}
This implies a higher density of particles around the equator, 
tapering off as particles flow along the shock. 

\begin{figure}
\resizebox{\hsize}{!}{\includegraphics[clip=true width=0.8\textwidth]{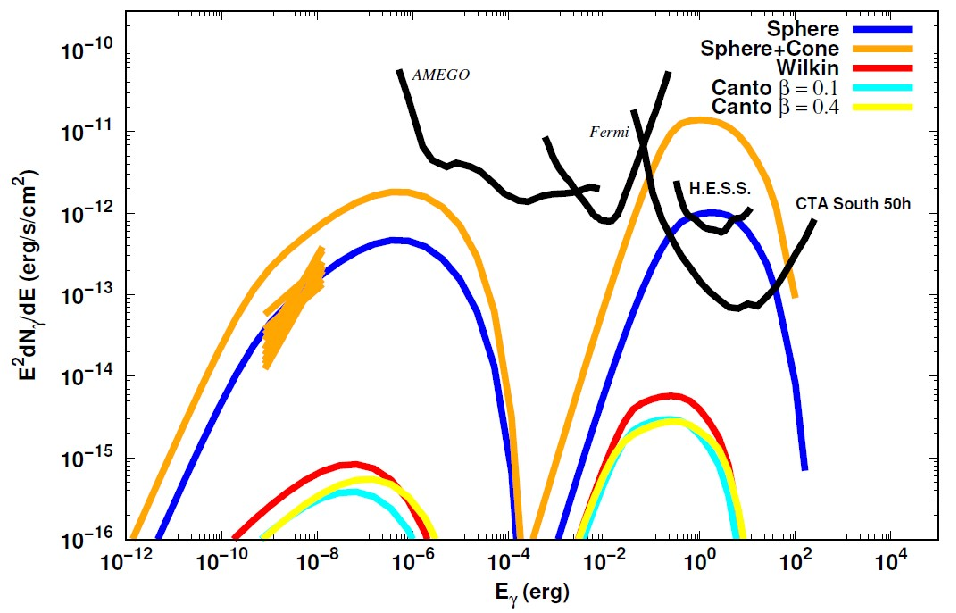}}
\caption{
\footnotesize
Sample SEDs at an orbital phase of $\phi_{\rm b}=180^\circ$ for several shock shapes in the case of the redback PSR J2339$-$0533 (with the shock wrapped around the pulsar). We used a millisecond pulsar mass of $M_{\rm NS}= 1.48M_\odot$, radius $R_{\rm NS}=
1.2\times 10^6$~cm, orbital period $P_{\rm b} =
4.63$~h, mass ratio $q =
4.61$, $R_0 = 0.25a$, $B$-field at the shock of $B_{\rm sh}=0.2$~G, companion surface temperature $T_{\rm comp }= 6.9\times 10^3$~K, $M_{\rm pair}=
500$, $\eta_{\rm p,max}=
0.8$, acceleration efficiency $\eta_{\rm acc}= 0.01$, power-law injection spectral index $\Gamma=1.7$, distance $d$ = 0.45~kpc, $\theta_{\rm max}=55^\circ$, maximum bulk flow $\beta\gamma_{\rm max}=7$, inclination angle $i=57^\circ$, and pulsar moment of inertia $I=1.7\times10^{45}$~g\,cm$^2$.
} 
\label{fig:SED_shocks}
\end{figure}

\begin{figure}
\resizebox{\hsize}{!}{\includegraphics[clip=true width=0.8\textwidth]{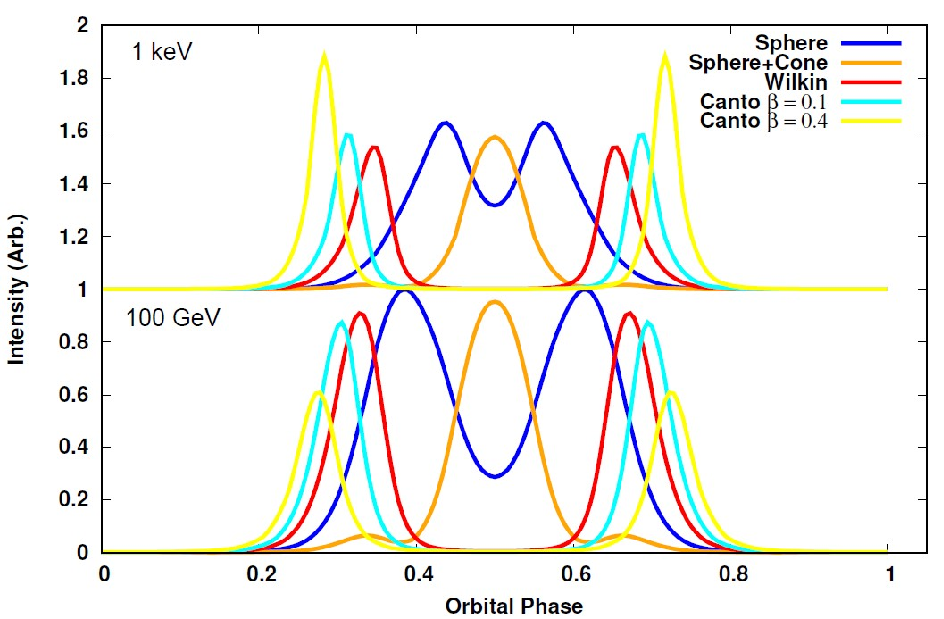}}
\caption{
\footnotesize
Same as Figure~\ref{fig:SED_shocks}, but for orbital light curves. The top panel is for 1~keV and the bottom panel for 100~GeV.
} 
\label{fig:LC_shocks}
\end{figure}

The effect of the different shock geometries on the broadband SED of PSR J2339$-$0533 is shown in Figure~\ref{fig:SED_shocks}. It is clear that the shock geometries change the angle of maximum intensity, since the velocity tangent changes. Thus, to recover the flux, one should probe different values of $i$. One can draw the same conclusion by considering the effect of the different shock geometries on the orbital light curves in Figure~\ref{fig:LC_shocks}.

\section{Conclusions}\label{sec:concl}
Spider binaries (especially nearby ones with hot and flaring companions) are promising targets for Cherenkov telescopes, exhibiting a rich multi-wavelength phenomenology. We are improving our \texttt{UMBRELA} code in several ways: using a smoother SR kernel, allowing more flexibility for the injection spectral shape, speeding up the code, interpolating spectra more carefully on the energy grids in the lab frame, and finally, studying the effect of different shock geometries on SEDs and orbital light curves. We find that the latter have a large impact on the observed flux for a given inclination angle $i$. In future, we will implement a skymap framework (projecting the emission on the sky for all $i$ and $\phi_{\rm b}$) to study the change in beaming pattern more closely. We are also planning to add more spectral components to the code, study radio eclipses and companion heating, as well as predicting polarisation signatures and assessing the contribution of spiders to the locally-measured positron excess above a few GeV.\\

\noindent {\bf{Authors}}\par
A.K.\, Harding$^{5}$,
M. Baring$^{6}$\\
\noindent {\bf{Affiliations}}\par
$^{5}$ Theoretical Division, Los Alamos National Laboratory, Los Alamos, NM 87545, USA\par
$^{6}$ Department of Physics and Astronomy - MS 108, Rice University, 6100 Main Street, Houston, Texas 77251-1892, USA.
 
\begin{acknowledgements}
This work is based on research supported wholly / in part by the National Research Foundation of South Africa (NRF; Grant Number 99072). C.V.\ acknowledges that opinions, findings and conclusions or recommendations expressed in any publication generated by the NRF-supported research is that of the author(s), and that the NRF accepts no liability whatsoever in this regard.
\end{acknowledgements}

\bibliographystyle{aa}
\bibliography{bibliography1}

\end{document}